\documentclass[10pt,journal]{IEEEtran}

\usepackage[pdftex]{graphicx}
\DeclareGraphicsExtensions{.pdf,.jpeg,.png}
\usepackage{graphicx}
\usepackage{multirow}
\usepackage{multirow,setspace,verbatim,amsfonts,graphicx,amsmath,amsthm,amsbsy,amssymb,epsfig}
\usepackage{url,cite}
\usepackage{booktabs}
\usepackage{comment}
\usepackage{amsmath}
\usepackage{breqn}
\usepackage{threeparttable}
\usepackage{array}
\usepackage{booktabs}
\usepackage{float}
\usepackage{hyperref}

\begin{document}

\title{AFP-CKSAAP: Prediction of Antifreeze Proteins Using Composition of k-Spaced Amino Acid Pairs with Deep Neural Network}

\author{Muhammad~Usman and Jeong~A.~Lee\\
	Chosun University, Korea\\
	Email: usman@chosun.kr, jalee@chosun.ac.kr
	\thanks{Copyright (c) 2019 IEEE. Personal use of this material is permitted.  Permission from IEEE must be obtained for all other uses, in any current or future media, including reprinting/republishing this material for advertising or promotional purposes, creating new collective works, for resale or redistribution to servers or lists, or reuse of any copyrighted component of this work in other works.}
	}
\maketitle
\begin{abstract}
Antifreeze proteins (AFPs) are the sub-set of ice binding proteins indispensable for the species living in extreme cold weather. These proteins bind to the ice crystals, hindering their growth into large ice lattice that could cause physical damage. There are variety of AFPs found in numerous organisms and due to the heterogeneous sequence characteristics, AFPs  are found to demonstrate a high degree of diversity, which makes their prediction a challenging task.  Herein,  we propose a machine learning framework to deal with this vigorous and diverse prediction problem  using the manifolding learning through composition of k-spaced amino acid pairs.  We propose to use the deep neural network with skipped connection and ReLU non-linearity to learn the non-linear mapping of protein sequence descriptor and class label.  The proposed antifreeze protein prediction method called AFP-CKSAAP has shown to outperform the contemporary methods, achieving excellent prediction scores on standard dataset. The main evaluater for the performance of the proposed method in this study is Youden's index whose high value is dependent on both sensitivity and specificity. In particular, AFP-CKSAAP yields a  Youden's index  value of 0.82 on the independent dataset, which is better than previous methods.
\end{abstract}


\IEEEpeerreviewmaketitle

\section{Introduction}
%
\IEEEPARstart{}Organisms living in the subzero temperature cope with the extreme cold environment using antifreeze proteins (AFPs). The temperature falling below the freezing point of water can cause the phenomenon of ice creation in the cells of the organisms which can be fatal \cite{kuramochi2019expression}. Also because of the property of re-crystallization of ice, the smaller crystals grows to become large ice lattice and cause physical damage to the cell membrane. AFPs hinders the ice crystal growth by binding onto the ice surface \cite{RAFP_original}. Antifreeze proteins are the heterogenous group of glycoproteins or  polypeptides  found in fish and other plants and animals \cite{iovine2019studying}. They are also responsible for the process of thermal hysteresis in which the freezing point is lowered without any alteration in the melting point of water, thus the gap between freezing and melting points is enlarged \cite{masud2019effect}. 

Two strategies that are observed for the cold hardiness adaptation or prevention  by the overwintering plants and animals are reported in \cite{sformo2009simultaneous}; the freeze tolerance in which the synthesis of ice nucleating agent is involved and the freeze avoidance in which ice-nucleating agents are removed in the species. To understand the process of interaction of AFP with ice, the utmost requirement is to be able to identify them reliably \cite{kandaswamy2011afp}. There is lack of similarity in the structure and sequences of the ice-binding proteins of even closely related species though they contribute to similar engagements, which in turn results in a complex prediction problem \cite{bhattacharya2018silico}. Antifreeze proteins have variety of applications in the food industry, biotechnological applications, medicine, preservation of organs and cell lines etc. Therefore, their reliable prediction is of utter importance \cite{pratiwi2017cryoprotect}.

Several researchers have successfully used machine learning approaches to address the classification of AFPs and non-AFPs. In  \cite{kandaswamy2011afp}, Kandaswany \textit{et.al.} proposed an anti-freeze prediction model also knwon as (AFP-Pred), which is considered as the first computational framework in this direction. They explored the usage of sequence information along with the physicochemical properties, short peptides and functional groups. Random forest approach was utilized for the identification of proteins.  In AFP\_PSSM \cite{xiaowei}, the evolutionary information of the AFPs were explored for classification with support vector machine (SVM). In AFP-PseAAC \cite{du2012pseaac},  using the features based on the Chou's pseudo amino acid composition, the information based on sequence order was explored as an additional feature. SVM classifier was used for the prediction of AFPs to achieve reasonable accuracy. In iAFP \cite{yu}, n-peptide composition of AFP from variety of organisms were analyzed using SVM and genetic algorithm. In particular, iAFP utilizes n=3 tripedtide composition features, however the experimental results are shown on very limited dataset. Another recently proposed method is afpCOOL\cite{afpcool}, in which four types of features were used to identify protein sequence through support vector machine. 
The most recent work in  this direction is  RAFP-Pred \cite{RAFP_original}, in which  protein sequence was equally divided into two segments and each segment is analysed for amino acid  and di-peptide composition features. The subset of significant features is selected using information gain and ranker method to design a random forest classifier. 

In this study we propose to use composition of $k$ spaced amino acid pairs (CKSAAP) to construct the feature vector for the prediction of AFPs. Application of CKSAAP can be found in variety of bioinformatics  \cite{chen2016profold},\cite{xu2018iglu},\cite{wei2014improved}. The encoding in this scheme is based on the frequency of amino acid pairs that are separated by $k$ number of residues.
In general, large values of $k$ contributes to  better prediction results \cite{zhao2012prediction}. However, increasing  the number of attributes eventually increases the computation complexity. The  conventional machine learning algorithms such as support vector machine or random forest can only generate competent prediction results when number of attributes are low \cite{singh2017impact}. On the other hand a well designed neural network can efficiently deal with large number of attributes for highly nonlinear classification tasks such as protein secondary structure prediction etc \cite{armenteros2019signalp}. Due to the exponentially large expressiveness, deep neural network model can easily learn highly non-linear mapping using ReLU and skipped connections \cite{yoon2018efficient}. 
To pre-process data for the deep neural network, a localized analysis is performed on the amino acid pairs using the notion of CKSAAP  to achieve a manifold learning approach for reliable prediction of AFPs.
Remainder of the paper is structured as follows: In section \ref{proposed}, the  proposed approach is discussed in detail. In section \ref{secresult}, the description of dataset, performance measures and evaluations are presented, which is followed by discussion in section \ref{discussion}. The  paper is finally concluded in section \ref{conclusion}.

\section{Proposed Approach}\label{proposed}
Machine learning approaches have been efficiently utilized in the area of protein prediction. The two vital steps to successfully address the prediction problem includes the formation of a significant feature set and selection of an appropriate machine learning algorithm. The later sections discuss the aforementioned issues.

\subsection{Dataset} \label{dataset1} 
For the evaluation of the suggested method in this study we use the standard dataset from \cite{kandaswamy2011afp}. The dataset is composed of protein sequences that are obtained from the Pfam database \cite{ sonnhammer1997pfam}. A stringent threshold was chosen (E=0.001) during the PSI-BLAST to remove any redundancy from the data. Furthermore, to retain only antifreeze proteins in the dataset, a manual search is performed. The final positive dataset contains the proteins which have less than 40\% similarity in the sequence. The non-antifreeze protein sequences in the dataset are obtained from seed proteins of Pfam family. The final dataset is composed of  9974 protein sequences  including 481 antifreeze proteins and 9493 non-antifreeze proteins. Two subsets are formed by  dividing dataset into $Tr$ for training and $Ts$ for testing.  $Tr$ contains a total of 600 protein sequences including 300  proteins from the AFP class and 300 proteins from non-AFP class. For generalization the sequences are selected randomly.  Testing was done on $Ts$ that is composed of leftover protein sequences from both  positive and negative datasets i.e 181 antifreeze proteins and 9193 sequences of non-antifreeze proteins.  We also consider the effect of imbalanced training dataset on validation performance, details of which is given in Section \ref{secresult}. 

\subsection{Features}\label{proposed1}
For feature construction, a sparse encoding scheme named CKSAAP is employed which has been used by researchers in various protein prediction problems \cite{chen2016profold} \cite{wei2014improved},\cite{chen2014towards}. CKSAAP stands for {\bf C}ompostion of {\bf k}- {\bf s}paced {\bf A}mino {\bf A}cid {\bf P}airs in which the frequency of the amino acid pair is calculated that are separated by $k$-residues ($k$ = 0,1,2,3,4,5...). This scheme basically represents the short and long range interactions amongst the residues along the sequence resulting in competitive performance in protein prediction problems \cite{chen2016profold},\cite{chen2011prediction}.
Selecting the value of $k$ = 0, CKSAAP encoding becomes identical to the di-peptide composition. The feature vector ($FV$) for $k$ = 0 can be defined as:
\begin{equation}
    FV = \left( \frac{F_{AA}}{N}, \frac{F_{AC}}{N}, \frac{F_{AD}}{N}, \dots, \frac{F_{YY}}{N} \right)_{400}
\end{equation}
\\where $N$ is the sequence length.\\

\begin{figure*}[ht]
    \centering 
    \includegraphics [width=16cm]{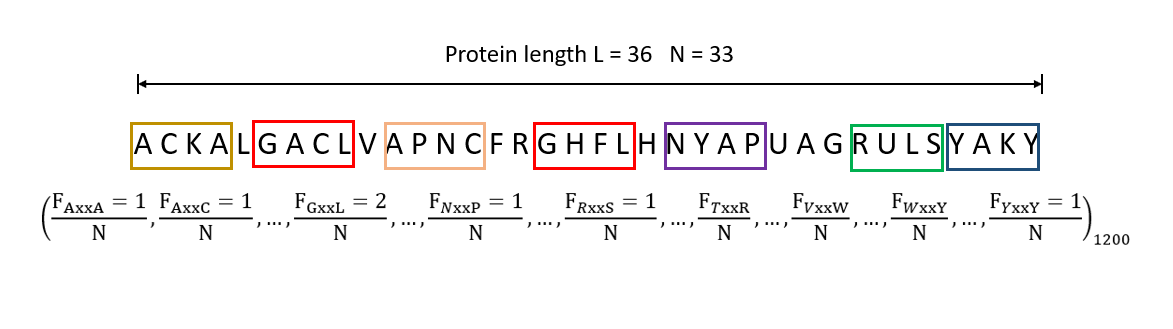}
    \caption{Illusration of CKSAAP descriptor calculation for $k=2$}
    \label{fig_comp}
\end{figure*}

The composition of residue pair in the protein sequence is represented by the value of each descriptor. For example if the residue pair $AC$ appears $j$ times, the composition of the corresponding residue pair is equal to $j/N$. For the protein length $L$, when $k$ = 0, 1, 2, 3, 4, 5 the values of $N$ is $L-1$, $L-2$, $L-3$, $L-4$, $L-5$ and $L-6$  respectively. Figure \ref{fig_comp} shows the encoding scheme for $k=2$.

\subsection{Classification}
Several machine learning based approaches have been proposed for the classification of proteins, for instance, random forest \cite{RAFP_original}, sparse representation classifier \cite{ecmsrc}, SVM \cite{afpcool} and multi-layer perceptron \cite{MLPhe2019prediction}. The classical machine learning algorithm show reasonable performance gain in most of the situations. However, conventional machine learning algorithms e.g., multi-layer perceptron neural network (MLP-NN) have inherited issues which hinder their scalability to higher layers. Due to the recent advancements in the deep learning theories \cite{ye2018deep}, we are now able to solve most of the issues with conventional MLP-NN. Therefore in this work we develop a deep neural network.
\\Deep learning has gained a lot of recognition in the area of machine learning in the recent era. We develop a four layered deep neural network architecture that is composed of an input layer, two hidden layers, and an output layer. The number of neurons in the input layer are equal to the feature's dimension while the first and second hidden layer has 70 and 128 neurons respectively. To avoid vanishing gradient issue in deep layers, we used rectified linear unit (ReLU) as the activation function and cross entropy as loss function. For better generalization, after every layer a dropout layer with $30\%$ is used. The idea of utilizing dropout is to arbitrarily drop some of the units alongwith the connections which in turns significantly reduces the overfitting issues \cite{srivastava2014dropout}. Additionally we use mini-batch gradient descent algorithm which selects a small portion of training sample at each iteration. Furthermore we utilize short and long skip connections to achieve faster convergence rate \cite{orhan2017skip}. The model is trained using Adam optimizer for 200 epochs.

\begin{figure}[ht]
	\begin{center}
		\centering
		\includegraphics[width=8cm]{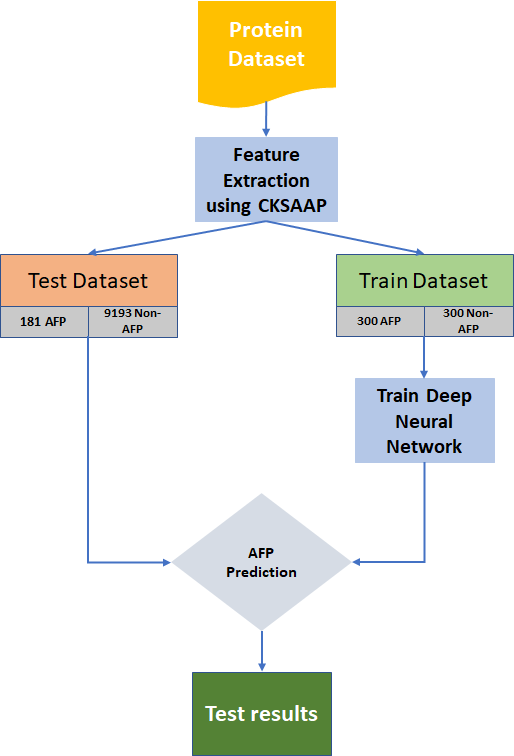}
	\end{center}
	\caption{Work-flow of the proposed approach.}
	\label{WorkFlow1}
\end{figure}

Figure \ref{WorkFlow1} shows the workflow of the proposed  classification model. Implementation of the model is done using Python on Keras (Tensor flow) and is freely available online for validation 

\section{Experimental Setup} \label{secresult}

\subsection{Performance measures}
\begin{figure}[ht]
	\begin{center}
		\centering
		\includegraphics[width=9cm]{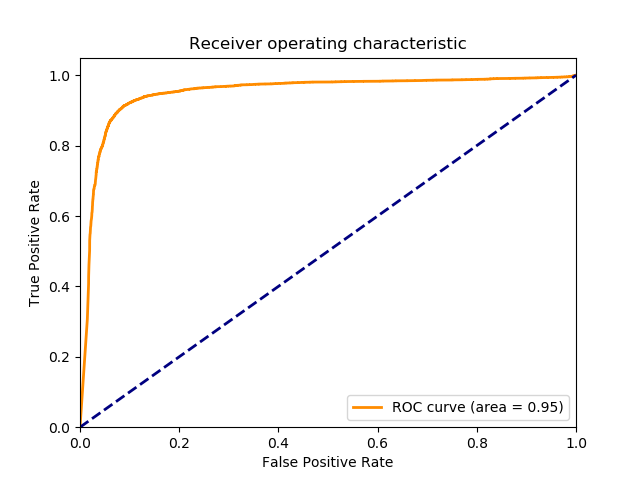}
	\end{center}
	\caption{ROC curve for the proposed algorithm}
	\label{roc}
\end{figure}

For the performance evaluation of the proposed framework we consider the widely used threshold-dependent biostatistics nomenclatures including sensitivity, specificity and accuracy. Furthermore Youden's index and Matthew's correlation coefficient (MCC) are  employed to measure the performance of the proposed method. The measurements can be defined using following equations:
\begin{equation}
    Sensitivity = \frac{TP}{TP+FN}
\end{equation}

\begin{equation}
Specificity = \frac{TN}{TN+FP}
\end{equation}

\begin{equation}
    Accuracy = \frac{TP+TN}{TP+TN+FP+FN}
\end{equation}

\begin{equation}
\resizebox{.45 \textwidth}{!} 
{
    $ MCC = \frac{TPTN-FPFN}{\sqrt{(TP+FP)(TN+FN)(TP+FN)(TN+FP)}}$
}
\end{equation}


\begin{equation}
    Youden's\! Index = Sensitivity + Specificity - 1
\end{equation}
Here TP, FP, TN and FN constitute to True Positive (correctly classified AFP), False Positive (incorrect classification of non-AFP as AFP), True Negative (correctly classified non-AFP) and False Negative (incorrect classification of AFP as non-AFP) respectively.
Therefore, specificity computes the fraction of non-AFPs that are correctly classified as non-AFP and sensitivity computes the fraction of AFPs being correctly classified as AFPs. The accuracy constitutes for the ratio of total number of correctly classified samples and the total number of samples. This indicates that in our case where the test dataset is highly imbalanced, a prediction model would yield a high accuracy value even if it classifies all the samples to the majority class which in our case is non-AFP. Therefore the prediction problems with imbalanced dataset should not be ranked according to the accuracy value they produce, instead they must be evaluated for the class specific performance measures such as Youden's index and MCC which are measured for joint performance gain.
Another comprehensive threshold-independent parameter is receiver operating characteristic (ROC). The ROC curve is plotted for the true positive rate (sensitivity) versus the false positive rate (1-specificity). Higher values of area under the curve (AUC) indicates better classification abilities of the predictor model. The ROC curve for the proposed prediction model is shown in Figure \ref{roc}.

\begin{figure}[ht]
	\begin{center}
		\centering
		\includegraphics[width=9cm]{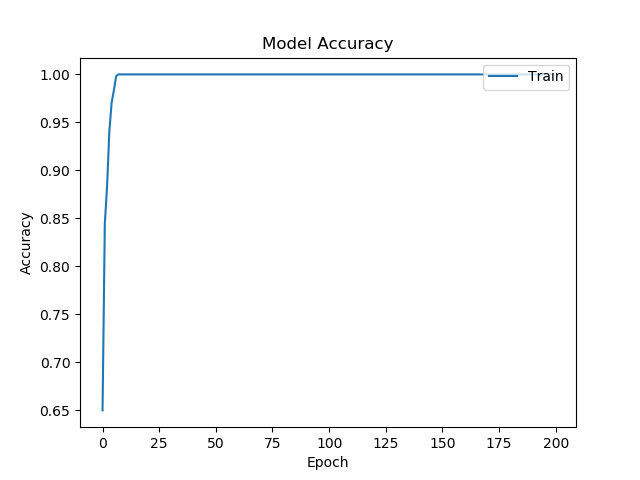}
	\end{center}
	\caption{Learning curve of the model}
	\label{Train_Curve}
\end{figure}

\subsection{Performance evaluation of the proposed AFP-CKSAAP}
Figure \ref{Train_Curve} shows the training curve of the algorithm over 200 epochs. The graph depicts that the model achieves 100\% accuracy in less than 10 iterations. To study the intervention of the value of $k$ on the performance, we evaluated the proposed classifier for different values of $k$ ranging from 0 to 13. The results reveal a significant increase in accuracy for the values between 0 to 8. However, values higher than 8 does not contribute to better performance for DNN. Therefore in proposed AFP-CKSAAP method we choose an optimum value of k = 8 for the prediction purpose. 

\begin{figure}[ht]
	\begin{center}
		\centering
		\includegraphics[width=9.5cm]{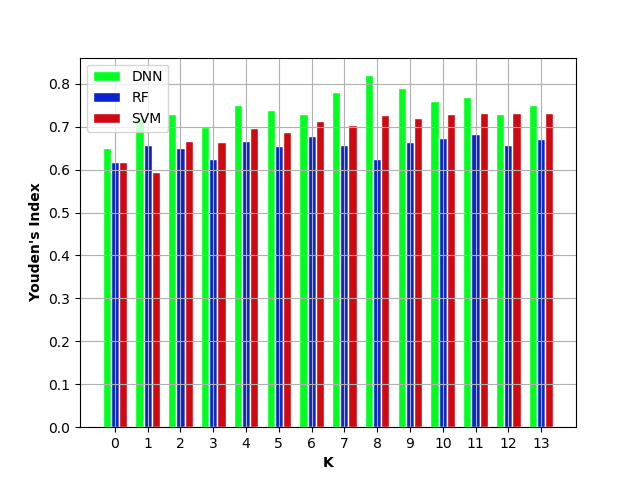}
	\end{center}
	\caption{Effect of varying values of (k) using DNN, Random Forest and SVM}
	\label{keffectall}
\end{figure}

\begin{figure}[ht]
	\begin{center}
		\centering
		\includegraphics[width=9.5cm]{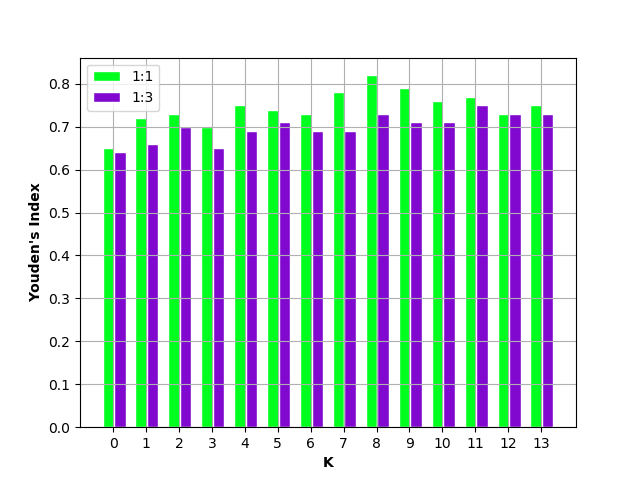}
	\end{center}
	\caption{Effect of varying amino acid pair gap (k) for distributed dataset using DNN}
	\label{keffect}
\end{figure}

For comprehensive performance evaluation of the proposed method, we  tested different values of $k$ using the balanced training dataset on random forest and SVM classifiers and compared with proposed classifier. In Figure  \ref{keffectall},  an improvement in the prediction's Youden's index with subsequent values of k for all the classifiers can be observed. Deep learning method however, outperforms random forest and SVM  approaches by a mean value of 0.10 and 0.07 respectively.  Furthermore to observe the effect of the distribution of dataset we evaluated the performance of the algorithm for imbalanced training datasets by selecting 3 times more negative samples i.e 300 AFPs and 900 non-AFPs as shown in Figure \ref{keffect}. The increase in the negative samples contributes to  increasing trend in the values of accuracy  as the classifier associates most of the samples to the majority class but the sensitivity measure is negatively affected. This proves that the accuracy should not be considered as an evaluation parameter for imbalanced datasets. Table \ref{testdata1_2} shows the performance comparison of the proposed classifier with the conventional machine learning classifiers. All the classifiers have been trained on the balanced dataset containing 300 sequences of both AFPs and non-AFPs using CKSAAP encoding scheme. Finally in Table \ref{testdata1_1}, the efficacy of the proposed method is compared with other machine learning methods in the literature. The results show that the proposed algorithm provides superior performance using the balanced training dataset.   

\begin{table*}[!ht]
	\begin{center}
		\caption{Comparison of several machine learning approaches on balanced training dataset using CKSAAP features with $k$ = 8}
		\begin{tabular}{lccccc}
			\hline
			{\bf Classifier} & {\bf Sensitivity } & {\bf Specificity } &  {\bf Accuracy} & {\bf Youden's index} & {\bf MCC}   \\
			\hline
			\\
			SVM (Linear)  &    80.1\% &    80.3\%  &    80.3\% &     0.60 & 0.20\\
			\hline
			Random Forest  &     76.8\% &    79.2\%  &    79.1\% &     0.56 & 0.18 \\ 
			\hline
			J48  &    76.8\% &    77.0\%  &    77.0\% &     0.53 & 0.17\\
			\hline
			IBk &    75.1\% &    74.0\%  &    74.0\% &     0.49 & 0.15\\
			\hline
			Attribute Selected Classifier (J48)  &    76.2\% &    69.6\%  &    69.8\% &     0.45 & 0.13\\
			\hline
			SVM (RBF)  &    69.6\% &    74.7\%  &    74.6\% &     0.44 & 0.14\\
			\hline
			Naive Bayes  &    62.3\% &    80.0\%  &    71.2\% &     0.42 & 0.43\\
			\hline
			BayesNet  &    68.5\% &    73.6\%  &    73.5\% &     0.42 & 0.13\\
			\hline
			{\bf Proposed (Deep Neural Network)} & {94.4\%} & { 87.8\%} & {88.0\%} & {0.82} & {0.33} \\
			\\
			\hline
		\end{tabular} 
		\label{testdata1_2}
	\end{center}
\end{table*}

\subsection{Comparison of AFP-CKSAAP with contemporary methods}\label{per_com}
In this section we compare the results of our proposed method with other methods reported previously. RAFP-Pred has been reported to produce good results in terms of Youden's index which is better than any previous approach. Table \ref{testdata1_1} shows that the proposed prediction method can predict the AFPs with the accuracy of 88.0\% which is better than AFP-Pred, AFP-PseAAC and comparable to CryoProtect. The algorithm demonstrates highest sensitivity value of 94.4\% which is almost 10\% greater than the next in order contributing in yielding the best value of 0.82 on the scale of Youden's index.  

\begin{table*}[!ht]
	\begin{center}
		\caption{Comparison of AFP-CKSAAP with contemporary approaches}
		\begin{tabular}{lcccccc}
			\hline
			{\bf Methods}&{\bf Classifier} & {\bf Accuracy} & {\bf Specificity} &  {\bf Sensitivity} & {\bf Youden's Index} & {\bf AUC}   \\
			\hline
			\\
			iAFP \cite{yu} & SVM &     95.6\% &    97.2\%  &    9.9\% &     0.07 & NA \\ 
			\hline
			AFP-Pred \cite{kandaswamy2011afp} & RF&   79.0\% &    79.0\%  &    82.3\% &     0.61 & 0.89\\
			\hline
			AFP\_PSSM \cite{xiaowei} & SVM &  93.0\% &    93.2\%  &    75.8\% &     0.69 & 0.93\\
			\hline
			AFP-PseAAC \cite{du2012pseaac}&SVM  &    87.5\% &    87.6\%  &    82.8\% &     0.70 & NA\\
			\hline
			afpCOOL\cite{afpcool} & SVM &96.0\% & 98.0\% &  72.0\% &  0.70 & NA \\
			\hline
			RAFP-Pred \cite{RAFP_original} & RF& 91.0\% & 91.0\% &  84.0\% &  0.75 & 0.95 \\
			\hline
			CryoProtect\cite{pratiwi2017cryoprotect} &RF & 88.2\% & 88.3\% &  87.2\% &  0.76 & NA \\
			\hline
			{\bf AFP-CKSAAP} & DNN& 88.0\% &  87.8\% &  94.4\% &  0.82 &  0.95 \\
			\\
			\hline
		\end{tabular} 
	  \begin{tablenotes}
	  \tiny
  \textit{iAFP features: n-peptide compositions}\\
  \textit{AFP-Pred features: Frequency of several functional groups + Physicochemical properties}\\
  \textit{AFP_PSSM features: AAC + DPC + PseAAC + Evolutionary information}\\
  \textit{AFP-PseAAC features: PseAAC}\\
  \textit{afpCOOL features: Hydropathy + Physicochemical + AAC + Evolutionary information}\\
  \textit{RAFP-Pred features: AAC + DPC}\\
  \textit{CryoProtect features: AAC + DPC}\\
  \textit{AFP-CKSAAP features: CKSAAP}
  \end{tablenotes}
		\label{testdata1_1}
	\end{center}
\end{table*}

\section{Discussion}\label{discussion}
In this study a predictor for antifreeze protein has been proposed that produced  better results. The classifier is composed of a deep learning neural network and the features are extracted using CKSAAP encoding scheme.  Many previous approaches have used binary encoding schemes to predict the AFPs. The binary encoding only represents the amino acids at distinct positions and loses the information of coupling of amino acids at different locations with in a sequence, whereas the CKSAAP encoder retrieves the information of amino acid correlation at different locations and also characterizes long and short interactions of amino acids within a sequence window. This information turns out to be extremely useful for achieving best prediction.

The conventional machine learning approaches used in the past does not show satisfactory performance when data with large number of attributes is to be processed. Previously, researchers have addressed the prediction problem using random forest and support vector machines. The deep neural network have gained immense popularity in the recent years and has been adopted in the fields of bioinformatics, computational biology, image and speech recognition. The deep layers can effectively extract complex functions and features from large data that can improve the classification of the prediction algorithms. 
\section{Conclusion}\label{conclusion}
Antifreeze proteins are responsible for the antifreeze effect of organisms living in subzero temperatures. The AFPs are useful in many applications of food industry and bioinformatics. Due to a great degree of diversity in the structure and sequences of  protein the classification of AFPs and non-AFPs is considered a challenging task.  Herein, we propose the use of deep neural network to serve the purpose of AFP prediction. The features are extracted from a sparse encoding scheme named CKSAAP which by varying window size describes the short and long range interactions of amino acid pairs in a sequence. Extensive experiments on the standard dataset have been performed.  The proposed AFP-CKSAAP approach exhibit outstanding performance on the scale of Youden's index compared to the prior predictors such as RAFP-Pred, iAFP, AFP\_PSSM, afpCool, AFP-Pred, CryoProtect and AFP-PseAAC. In our future work we aspire to come up with a systematic strategy for the calibration of the neural network and also to explore the structural information of the proteins. The Python implementation of AFP-CKSAAP can be downloaded from 

\section{Acknowledgments}
This research was supported by Basic Science Research Program through the National Research Foundation of Korea (NRF), funded by the Ministry of Education (No. 2019R1I1A3A01058887) and in part by Korea Institute of Energy Technology Evaluation and Planning, and the Ministry of Trade, Industry and Energy of the Republic of Korea (No. 20184010201650).

\bibliographystyle{IEEEtran}
\bibliography{AFP-CKSAAP}

\end{document}